\documentstyle[11pt]{article}
\input{epsf} 

\newcommand{\epscfig}[3]{\begin{figure}[htb]\begin{center}\epsfysize #2\
\epsfbox{#1} \newline \protect #3\end{center}\end{figure}}

\title{Nuclear liquid-gas phase transition within the lattice gas model}
\author{J. Borg$^a$, I.N. Mishustin$^{a,b}$, J.P. Bondorf$^a$ \\
{\it $^a$ The Niels Bohr Instiute, Blegdamsvej 17, DK-2100 Copenhagen \O,
  Denmark} \\
{\it $^b$ The Kurchatov Institute, Russian Research Center, 123182 Moscos, Russia}}

\begin{document}
\maketitle
\begin{abstract}
We study the nuclear liquid-gas phase transition on the basis of a two-component lattice gas model. A Metropolis type of sampling method
is used to generate microscopic states in the canonical ensemble. The
effective equation of state and fragment mass distributions are evaluated
in a wide range of temperatures and densities. A definition of the phase
coexistence region appropriate for mesoscopic systems is proposed. The caloric
curve resulting from different types of freeze-out conditions are presented. 
\end{abstract}

\vspace{2cm}
It is commonly believed that nuclear matter undergoes
a liquid-gas phase transition at lower densities and temperatures.
In the early 1980s it was suggested \cite{siemens83} that this 
phase transition might be probed in heavy-ion collisions at intermediate
energies by observing the disintegration of colliding nuclei into 
many fragments of different sizes, a phenomenon commonly known
as {\it multifragmentation}. 

A wide variety of models has been proposed
for nuclear multifragmentation, ranging from statistical to dynamical ones
(see e.g. \cite{bondorf95} for a review).  
In principle, most of these models yield an equation of state of nuclear matter. The evaluation
of the equation of state is, however, in general rather difficult and 
has only been done in some simple cases \cite{csernai94}.
In this paper we focus on the relationship between the fragmentation
of a nuclear system and the corresponding equation of state, calculated
on the basis of a two-component {\it lattice gas model}. In particular,
we will discuss the nuclear {\it caloric curve} (temperature versus
excitation energy) resulting from different types of freeze-out conditions.
The details of the model and the numerical simulations will be further
discussed in a forthcoming article \cite{new_article}. Here, we only present
some main features and results of the model.

Common for all types of lattice gas models is the use of a spatial lattice,
with a certain dimension, number of sites, $N_V$, and a certain 
Bravais-structure, such that each lattice point has a fixed number
of neighbours $\gamma$. For simplicity and for comparison with
earlier works we consider here a simple cubic $(s.c.)$ lattice with $\gamma=6$. 
The lattice spacing $a_0$ is chosen such that the density of 
a fully occupied lattice equals the nuclear ground state density 
$\rho_0=0.16~\mbox{fm}^{-3}$, i.e. $a_0^3=1/\rho_0$.
The effective nucleon-nucleon $(NN)$-interaction is represented
by a nearest-neighbour square well potential $V$ with a repulsive hard core.
In terms of the discritized lattice distance $r$ between the nucleons 
$(r=0,a_0,\sqrt{2}a_0,\ldots)$ the potential is given as,
$$
V_{ij}(r)=\left\{ \begin{array}{ll} \infty & \mbox{for $r=0$} \\
\epsilon_{{t_3}_i{t_3}_j} & \mbox{for $r=a_0$} \\
0 & \mbox{for $r>a_0$}\end{array}\right.
$$
Here, $t_{3_i}$ denotes isospin of nucleon $i$ and 
$\epsilon_{{t_3}_i{t_3}_j}$ are the interaction strength parameters.
In fact we only allow for two such parameters, $\epsilon_s$ for 
$pp$ and $nn$ interacting pairs and $\epsilon_d$ for $pn$ pairs.
According to the dynamical potentials and previous works on lattice models
\cite{gupta} $\epsilon_d$ is chosen slightly repulsive,
$\epsilon_d=1~\mbox{MeV}$. The other parameter, $\epsilon_s$, is chosen such
that the nuclear binding energies are reproduced, $\epsilon_s=-5~\mbox{MeV}$.

If $\vec{p_i},\vec{r_i}$ denote respectively the momentum and position of
nucleon $i$, the Hamiltonian of the system of $A$ nucleons can be written as,
\begin{equation}
{\cal H}={\cal T}+{\cal V}=\sum_i^A \frac{\vec{p_i}^2}{2m}+
\sum_{i<j} V_{ij}(\vec{r_{ij}}),
\end{equation}
where $m$ is the nucleon mass. It can be shown that the interaction
Hamiltonian ${\cal V}$ is homomorphic to the {\it spin-1 Ising model}, which is well-known in
solid state physics. If $\epsilon_s$ is put equal to $\epsilon_d$ the model reduces
to the standard lattice gas model which is isomorphic to
the spin-1/2 Ising model. This model was first considered in context of
nuclear fragmentation in \cite{biro86} and later used by S. Das Gupta and
J. Pan \cite{gupta,gupta_mean,gupta_sim} and
X. Campi and H. Krivine \cite{campi97}. The two-component lattice gas model has
also been used in \cite{gupta_sim,gupta_mean}.

In the following we will for simplicity treat the model in a canonical
ensemble. This implies, that the temperature
$T$, the volume $V$, the number of particles $A$ and the asymmetry of the
system, $y=Z/A$, are the control variables. In this paper we
only consider symmetric matter, $y=0.5$.

The states $f$ of the nuclear system can be classified by the positions, momenta and isospins of all the nucleons,
$f=\{ (\vec{r},\vec{p},t_3)_i | i=1,..A\}$. Due to the classical nature of
the model the canonical partition function, $Z$, can be factorized into two
parts coming respectively from the momentum space and the configuration space.
The first part is
proportional to the partition function of an ideal gas.
The contribution of the momenta of the
nucleons to the various thermodynamical quantities can therefore be
calculated in a straightforward way.

The partition function and thus the Helmholtz free
energy, $F$, of the configuration space is far less trivial to obtain. 
In this work it is done numerically by performing a Monte Carlo type of
sampling.  The realization of this sampling method consists of three parts: 
the initialization, the equilibration of the system and the calculation of ensemble averages.

The initialization of the system is done by putting the particles one by one
in the lattice. At each step the spatial probability distribution is changed
with a Boltzmann factor, $e^{-\epsilon_{ij}/T}$, in the neighbourhood of the
first particle. This method of initializing the
system is identical to the one used in \cite{gupta}.

The equilibration of the system is performed 
by constructing a random walk of microscopic states $\{f_{\nu}\}$ through the 
configuration space via a {\it Markov process}\footnote{A sequence of states
each of which depends only on the preceding one}, starting from the
initial state, $f_{\nu=0}$. The Markov process is
defined by specifying a general transition probability matrix, 
$p(f\rightarrow f')$ from the state $f$ to the state $f'$. The choice of the
transition matrix is dictated by the detailed balance condition. In the
present work it is defined as,
\begin{equation}
\label{transition}
p(f\rightarrow f')=\left\{ \begin{array}{ll} 
e^{-\beta\Delta {\cal V}} & \mbox{if $\Delta {\cal V} > 0$} \\
1 & \mbox{otherwise}\end{array}\right.
\end{equation}
Here, the test function, $f'$, is obtained from $f$ either by changing the position of a randomly
chosen nucleon or by interchanging a randomly chosen proton-neutron pair.
In case of acceptance, the trial function $f'$ will be the new state of the
system $f_{\nu+1}=f'$. Otherwise, the system remains in the same state,
$f_{\nu+1}=f_{\nu}$.  It can be proven formally \cite{alben77} that the
above choice of $p$ leads to the desired convergence property, i.e. 
the equilibrated probability distribution of microscopic states will
correspond to a macroscopic state of minimum free energy.
This algorithm is similar to the one introduced by Metropolis et. al \cite{metro}.

When equilibrium has been reached, the ensemble average $<\! Q \!>$ of any
quantity $Q$ can be calculated by repeating the same procedure a
certain number of times and recording the value of the quantity, $Q_{\nu}$,
at each time step $\nu$.

By a systematic variation of the density $\rho=A/V$ and the temperature $T$
the algorithm one can eventually 
probe all the different points of the phase diagram including the coexistence
region. 
\footnote{This is an important difference to the grand canonical ensemble
which only allows the system to be in a single phase.}
The density is varied by varying the number of particles in the lattice
for a {\it fixed} volume. The volume is put to $N_V=6^3$ which approximately
corresponds to the maximum fragment mass, $A_{max}=190$, observed in
heavy ion experiments. The boundaries of this lattice are chosen to be open
to mimic the effect of the vacuum surrounding the real nuclear
system. The results for periodic boundary conditions will be presented
elsewhere \cite{new_article}.

The quantities, $Q$, of interest are the ensemble average 
$E(T,\rho)=<\!{\cal V}\!>$ of the potential energy ${\cal V}$, and the 
distributions of cluster sizes,
i.e. the mean number of fragments, or the yield $Y(A)$ as a function of the
mass number $A$. Contrary to most statistical models of
nuclear fragmentation (see i.e. \cite{bondorf95} and references herein), clusters are 
not explicitly defined in the lattice gas model. 
In our simulations we implement a definition based on pairwise binding. Two neighbouring nucleons are thus prescribed to the same cluster if their relative kinetic energy is not
sufficient to overcome the attractive bond:
$\vec{q}^2/4m+\epsilon_{{t_3}_i{t_3}_j}<0$, where
$\vec{q}=\vec{p}_1-\vec{p}_2$ is the relative momentum between
the two particles. This definition was originally proposed by Hill \cite{hill}
and has also been used by S. Das Gupta et. al \cite{gupta}. 

By applying the standard laws of thermodynamics all other quantities can be 
derived from the internal energy $E$ of the system in the different 
points, $\rho,~T $, of the phase diagram.
One of the most interesting applications is the study
of the liquid-gas phase transition in a finite system. The first order 
phase transition
is signalled by the presence of the phase co-existence region. In a finite
system the definition of this region is somewhat ambiguous. In the
calculations below such regions are identified by the existence of 
two distinct densities, $\rho_g<\rho_l$, for which the following conditions are fulfilled,
\begin{equation}
\label{binodal}
\mu_l(\rho_l,T)=\mu_g(\rho_g,T)=\mu ,~~~F(\rho_l,T)/V=\mu(\rho_l-\rho_g)+F(\rho_g,T)/V
\end{equation}
The points $(\rho_g,T)$ and $(\rho_l,T)$ belong to respectively the gas
branch and the liquid branch of the co-existence curve. Geometrically, the above criteria
correspond to the requirement that the points $(\rho_g,f_g)$ and 
$(\rho_l,f_l)$ belong to a {\it common tangent} of the free energy density,
$f=F/V$.
For a macroscopic system the free energy density would follow
a straight line in the region $\rho_g \le \rho \le \rho_l$. For a mesoscopic
system the free energy density is slightly larger in this region due to the
presence of an interface between the two phases. 

The equation of state for the system with volume $N_V=6^3$ is depicted in Fig.
1 with the pressure $P$ as a function of density $\rho/\rho_0$ and
temperature $T$. The thick solid line is the co-existence curve calculated
according to the definition (\ref{binodal}). The two branches of the curve
terminate at the critical point 
$(\rho/\rho_0,T,p)_{cr}\approx (0.39,4.8~\mbox{MeV},0.15~\mbox{MeV/fm$^3$})$.
For comparison, the critical point in the macroscopic limit is
found to be 
$(\rho/\rho_0,T,P)_{cr}\approx (0.58,6.7~\mbox{MeV},0.36~\mbox{MeV/fm$^3$})$ 
\cite{new_article}.
In the co-existence region for a fixed temperature the pressure as a function
of the density  is almost a constant.

One should bear in mind that in experiment one observes only a final state of
the system where the fragments cease to interact (freeze-out configuration).
It is not clear a priori which thermodynamical conditions correspond to these
configurations. Below we consider two possibilities corresponding to constant
pressure and constant density at freeze-out. The two lines across the co-existence region are defined by $p=0.10~
\mbox{MeV/fm$^3$}$  and $\rho/\rho_0=0.3$, respectively. The letters $A-E$ 
along the isobar are chosen with a fixed step in densities 
$\Delta\rho/\rho_0=-0.15$ starting from $\rho/\rho_0=0.75$. the letters
along the isochor are chosen at the temperatures 
$T=(2.4,3.5,4.3,4.9,6.5)~\mbox{MeV}$ (points $c$ and $D$ coincide).

The cluster size distribution corresponding to these points of the phase
diagram are shown in fig. 2a and 2b for the isobar $p=0.10
~\mbox{MeV/fm$^3$}$ and the isochor $\rho/\rho_0=0.3$ respectively. 
At point $A$ the system essentially consists of a compound nucleus surrounded
by nucleons and small clusters. Point $A$ belongs to the borderline
of the co-existence region. At point $B$ the compound nucleus 
starts to break-up producing 
some intermediate mass fragments (IMFs: $4\le A\le 20$). At point $C$ and $D$ fragments 
of all different sizes are present, whereas at point $E$ the size distribution
is exponentially decreasing indicating that only the gas phase is present.
Thus the appearance of IMFs is a clear manifestation of the first-order
liquid-gas phase transition. 

The evolution of the size distribution is observed to be smoother 
along the isochor. Point $a$ belongs to a transition region for which
the pressure increases rather abruptly. At even lower temperatures
the system essentially consists of one compound nucleus surrounded by vacuum
(not shown). The sudden increase of the pressure results from the break-up of
this nucleus. As can be seen in the plot this produces some IMFs.
The distribution at point $d$ still contains a significant contribution of
IMFs although it does not belong to
the co-existence region, according to definition of eq. (\ref{binodal}). 
Thus, the cluster size distribution alone can not give a unique
relation between multifragmentation and liquid-gas phase transition. 
At point $e$ the distribution is closer to the exponential one.

Figures 2a and 2b indicate that cluster size distributions are widest at the 
vicinity of the critical point. If the cluster size distribution
is approximated by a power law $Y(A)\sim A^{-\tau}$, 
one finds that a good fit can be achieved for a wide range of temperatures
and densities around the critical point. The range of fragment masses used for the power law fit was chosen to $1\le
A\le 20$, but the conclusions are not very sensitive to this choice. Typical
values of $\chi^2$ per degree of freedom are in the range of 
$0.03-0.13$. Our analysis shows that
for each isochor, $\tau$ has a minimum 
at the binodal line. Furthermore, $\tau$ has a global minimum point in the vicinity of
the critical point, where $\tau \approx 2.2 $. This is in accordance with 
the value of the Fisher exponent for the Ising model,
$\tau=2.21$. 

From the yield of clusters $Y(A)$ it is not easy to distinguish between
the two types of freeze-out conditions. The difference reveals itself
more clearly in the associated caloric curves, showing the variation of the
temperature with respect to the 
excitation energy $E^*$ per nucleon. Here, the excitation energy is defined as
$$
E^*(A,T)=3/2\cdot T\cdot(<\!M\!>-1)+E(A,T)-E(A,T=0),
$$
where $<\!M\!>=\sum_A Y(A)$ is the total multiplicity of the system. 
Note, that we identify the kinetic part of the excitation energy 
with the total kinetic energies of the clusters rather than the total
kinetic energy ${\cal T}$ of the individual constituents. 
The caloric curves for 
$p=0.1~\mbox{MeV/fm$^3$}$ and for $\rho/\rho_0=0.3$ are
shown in Fig. 3. It is seen that the pronounced plateau in the caloric curve appears
only in case of the fixed pressure condition.
In both cases the caloric curve
approaches a straight line for high temperatures. At low temperatures
the model predictions are not accurate, since quantal effects cannot be
neglected in this region. Qualitatively, the caloric curve at constant
pressure looks very similar to the one predicted by the Statistical
Multifragmentation Model \cite{SMM} and recently observed in the ALADIN experiment \cite{aladin}. 

To each of curves on the phase diagram, which correspond to a specific change
of the system state in a reversible process, a particular heat capacity $C$
can be attributed. Our analysis shows, that the specific heat capacity for 
fixed pressure, $C_p$, in fact has a {\it singularity} along subcritical
isobars. This infinity is related to the latent heat of the
transition. On the other hand, due to finite size effects the specific heat 
capacity for fixed number of particles and volume, $C_V$, is finite
in all points of the phase diagram. We will elaborate further on these points
in a forthcoming publication \cite{new_article}.

In conclusion, our results demonstrate the close connection
between the nuclear liquid-gas phase transition and the nuclear
multifragmentation. The calculations have been made on the basis of a
two-component lattice gas model.  A sampling method has been presented by
which both the effective equation of state and the fragment mass
distributions have been evaluated in a consistent way. The method has an advantage compared to other statistical
approaches to nuclear multifragmentation, where the link between 
the cluster observables and the corresponding thermodynamical
properties of the nuclear system is less transparent.

We have proposed a definition of the phase transition, more 
relevant for finite systems, which is based on the identification of an interphase
between the two phases. The relevance of this definition is supported by
the critical exponent analysis and the singular behaviour of the specific
heat capacity, $C_p$, along subcritical isobars.
We have examined the caloric curve resulting from two different
freeze-out conditions. It was demonstrated that the occurrence of a plateau
in the caloric curve seems to be consistent with the condition of constant
and subcritical pressure in the freeze-out configurations. 

There are of course some obvious deficiencies of the model. 
First, the true Fermi-statistical nature of the nucleons can not 
be neglected for lower excitation energies. 
Secondly, the idea of a thermal bath is certainly an idealization and it is necessary
to perform a real microcanonical sampling in the Monte Carlo
algorithm. Third, for larger systems it is necessary to include the
Coulomb repulsion and determine the corresponding change of the phase
diagram. Finally, the prescription of clusters in the model is still not
unambiguous. The clusters defined by the pairwise binding are generally
excited and should undergo de-excitation at later stages (see \cite{barz}).
In the future, we plan to improve the model in various respects to take
these deficiencies into account.

\section*{Acknowledgements} 
We thank S. Das Gupta, A.D. Jackson and J. Pan for
stimulating discussions. One of us (I.N.M) acknowledges the financial support
from the Carlsberg Foundation (Denmark).

\newpage
\section*{Figure captions}
{\bf Fig. 1}. Pressure $P$ as a function of density
$\rho/\rho_0$ and temperature $T$ for a lattice with $N_V=6^3$ sites. The
thick solid line is the co-existence curve. The line with capital letters
is the isobar $p=0.1~\mbox{MeV/fm$^3$}$. The line with small letters
is the isochor $\rho/\rho_0=0.3$.

\vspace{1cm}
{\bf Fig. 2}. The yields $Y(A)$ as functions of the fragment mass 
$A$ corresponding to different points of the phase diagram: {\bf a)} The
points $(A-E)$ belong to the isobar $p=0.10~\mbox{MeV/fm$^3$}$, {\bf b)} the
points $(a-e)$ belong to the isochor $\rho/\rho_0=0.3$. Note the different
scales on the x-axis.

\vspace{1cm}
{\bf Fig. 3}. The temperature $T$ as a function of the specific excitation
energy $E^*/A$ for two different freeze-out conditions. The solid line
corresponds to the fixed pressure condition, $p=0.1~\mbox{MeV/fm$^3$}$, and
the dotted line corresponds to the fixed density condition,
$\rho/\rho_0=0.3$. The letters along the lines refer to fig. 1.

\newpage 

\epscfig{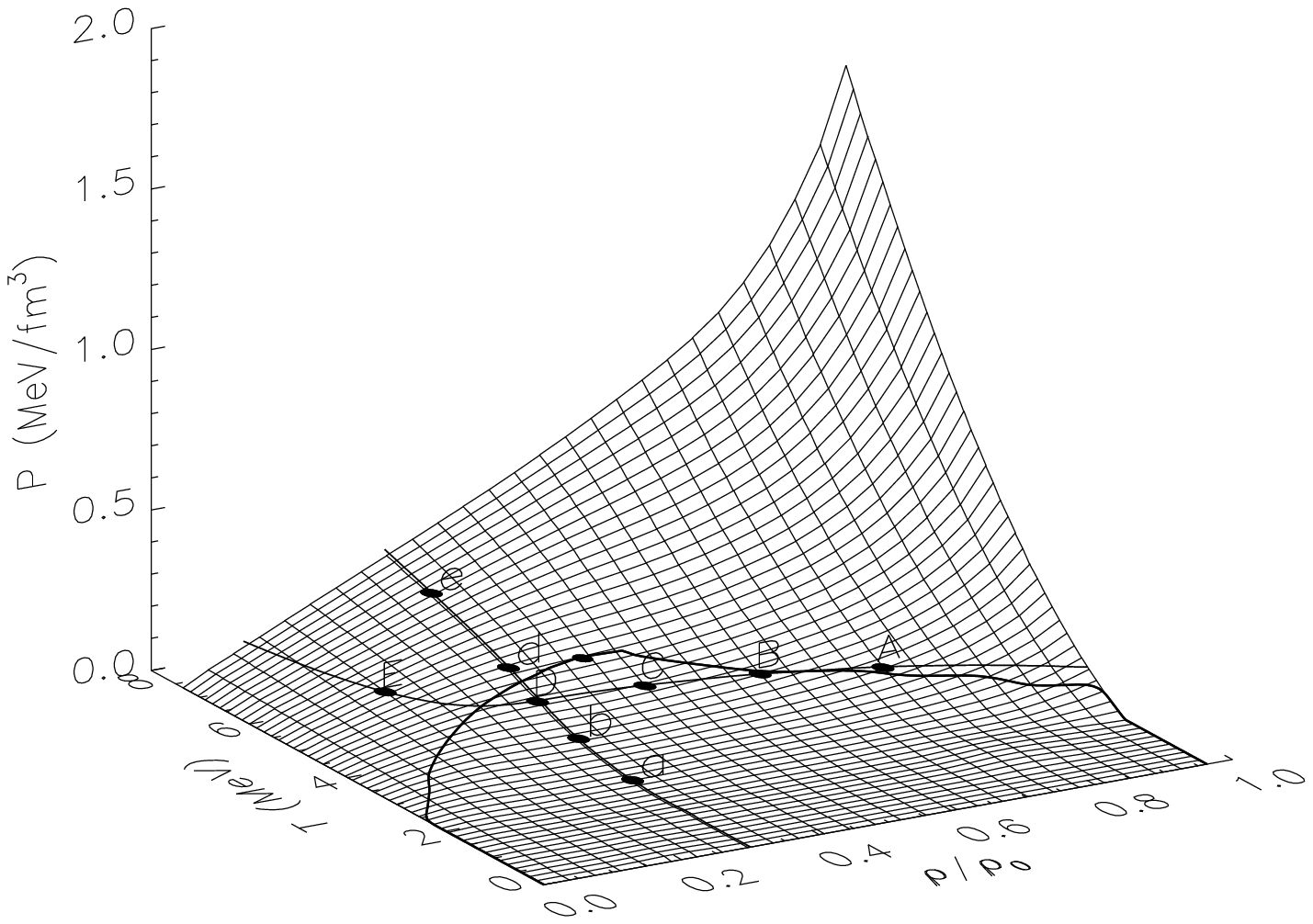}{11cm}{Figure 1}

\epscfig{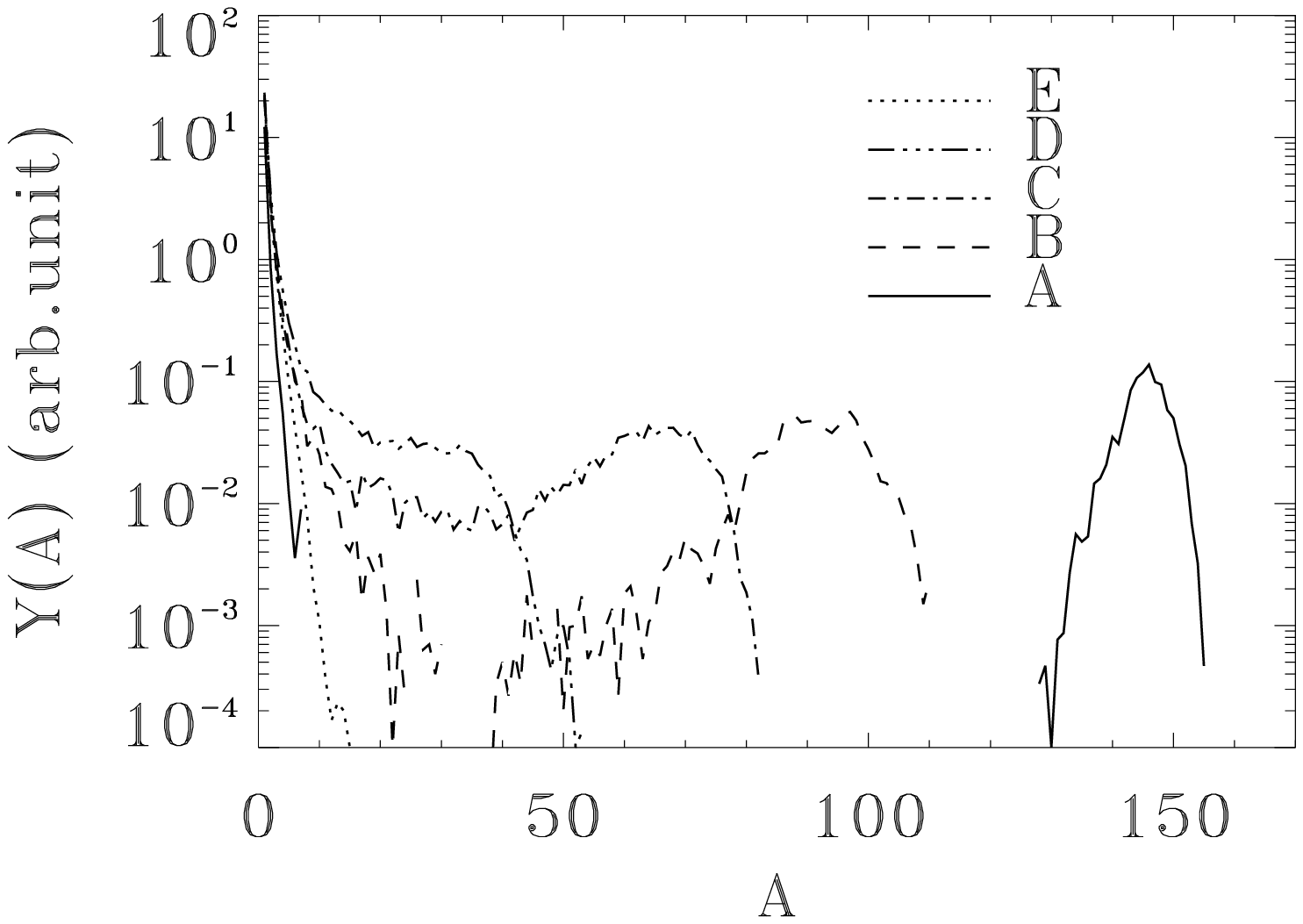}{7.5cm}{Figure 2a}

\epscfig{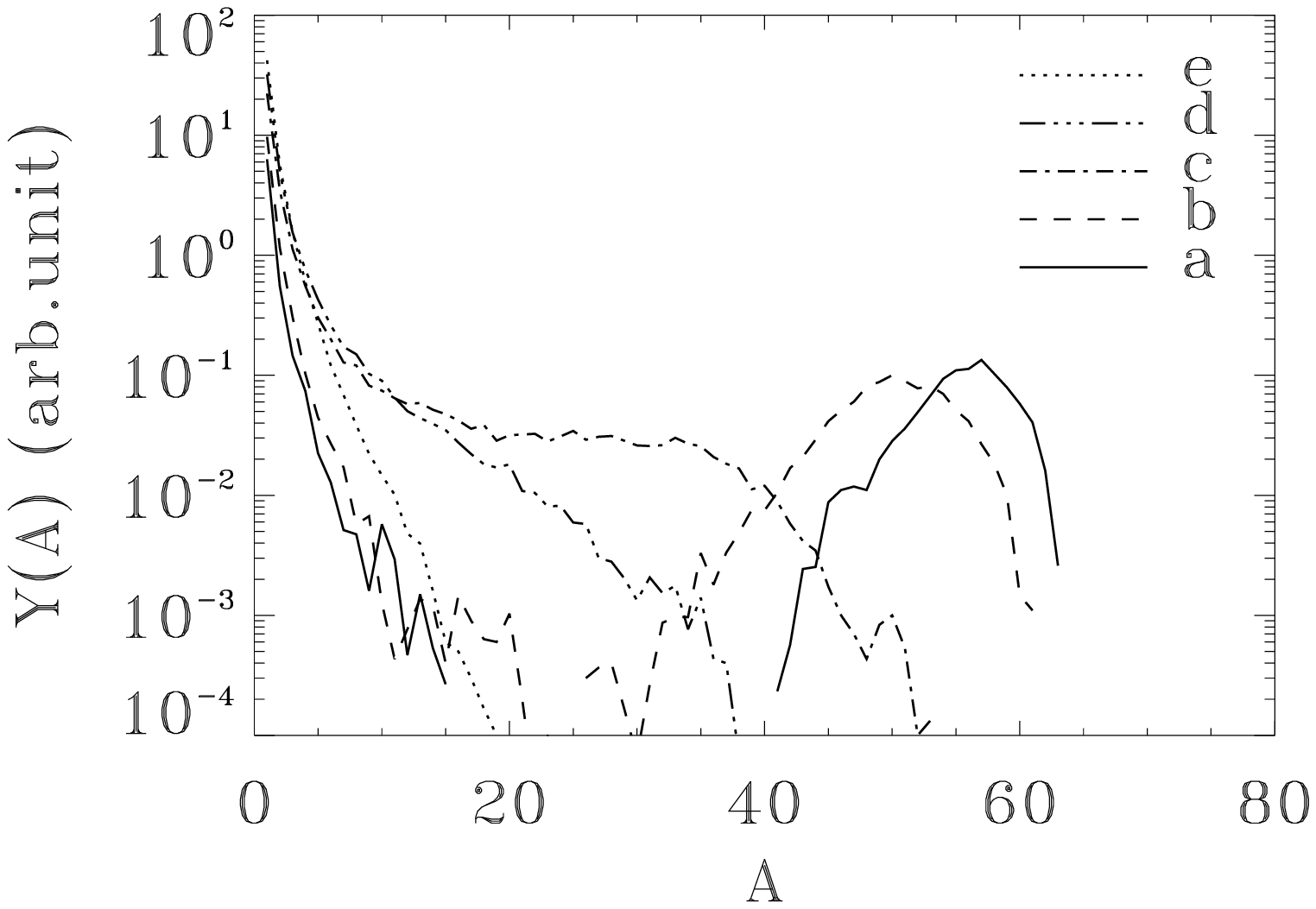}{7.5cm}{Figure 2b}

\epscfig{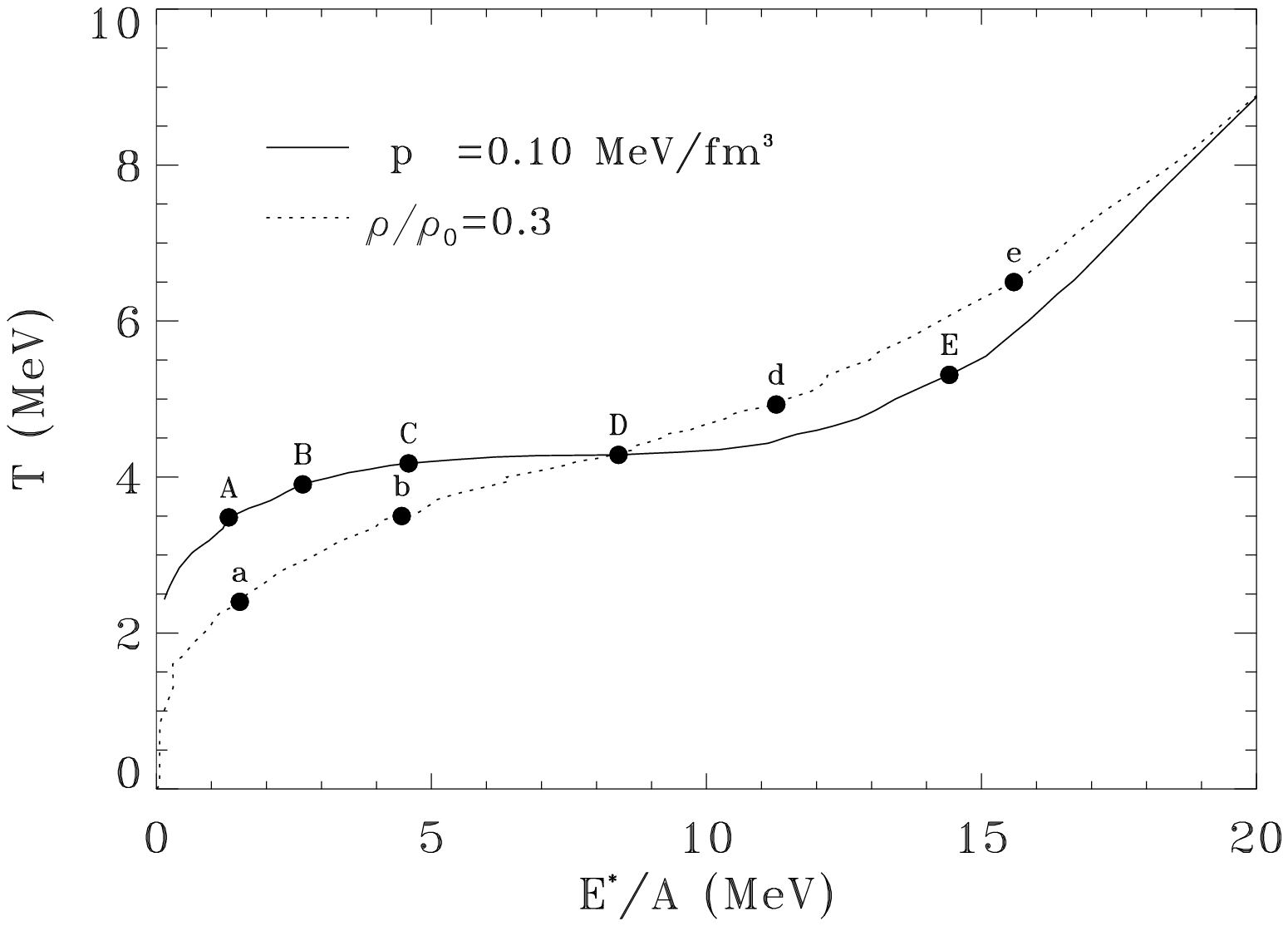}{11cm}{Figure 3}


\newpage
\begin{thebibliography}{99}
\bibitem{siemens83}
P.J. Siemens, \emph{Nature} {\bf 305} (1983) 410. \newline
G. Bertz, P.J. Siemens, \emph{Phys. Lett. B} {\bf 126} (1983) 9.
\bibitem{bondorf95}
J.P. Bondorf, A.S. Botvina, I.N. Mishustin, A.S. Iljinov,
  K. Sneppen,\emph{Phys. Rep.} {\bf 257} (1995),133-219
\bibitem{csernai94}
L. P. Csernai, \emph{'Introduction to Relativistic Heavy
    Ion Physics'} (1994), ed. John Wiley \& Sons Ltd, (Chichester, West Sussex PO19 1UD, England)
\bibitem{new_article}
J. Borg, I.N. Mishustin, \emph{Thermodynamical properties of
  finite nuclear systems within the lattice gas models}, in preparation.
\bibitem{gupta}
S. Das Gupta, J. Pan, \emph{Phys. Lett. B} {\bf 344} (1995) 29. \emph{Phys. Rev. C} {\bf 51} (1995) 1384. \emph{Phys. Rev. C} {\bf 53} (1996) 1319. \emph{Phys. Rev. C} {\bf 54} (1996) R2820. \emph{Phys. Rev. Lett.} {\bf 80} (1998) 1182. 
\bibitem{biro86} 
T.S. Biro, J. Knoll, J.Richert,
\emph{Nucl. Phys.} {\bf 1459} (1986) 692.
\bibitem{gupta_sim}
S. Das Gupta, J. Pan, I. Kvasnikova, C. Gale, \emph{Nucl. Phys. A} {\bf 621}
(1997) 897.
\bibitem{gupta_mean} 
J. Pan, S. Das Gupta \emph{Phys.Rev. C} {\bf 57} (1998) 1839.
\bibitem{campi97} 
X.Campi, H. Krivine, \emph{Nucl. Phys. A}, {\bf 620} (1997) 46.
\bibitem{alben77}
R. Alben, S. Kirkpatrick and D. Beeman, \emph{Phys. Rev. B} {\bf 15} (1977)
346.
\bibitem{metro}
N. Metropolis, A.W. Rosenbluth, M.N. Rosenbluth, A.H. Teller, E.Teller,
\emph{J. Chem. Phys} {\bf 21} (1953) 1087.
\bibitem{hill}
T.L. Hill, \emph{J. Chem. Phys.} {\bf 23} (1995) 617.
\bibitem{SMM}
J. Bondorf, R. Donangelo, I.N. Mishustin and H. Schulz, \emph{Nucl. Phys. A}
{\bf 444} (1985) 460.
\bibitem{aladin}
J. Pochodzalla and the ALADIN collaboration, \emph{Phys. Rev. Lett.} {\bf 75}
(1995) 1040.
\bibitem{barz}
H.W. Barz, J.P. Bondorf, D. Idier, I.N. Mishustin, \emph{Phys. Lett. B} {\bf
  382} (1996) 343.
\end{thebibliography}
\end{document}